\documentclass[aps,  prb,twocolumn,
  superscriptaddress,  floatfix,  10pt         
]{revtex4-2}
\usepackage{float}
\usepackage{graphicx}
\usepackage{amsmath,amssymb}
\usepackage{tikz}
\usepackage{braket}
\usepackage{pgfplots}
\pgfplotsset{compat=1.17}
\usepackage[version=4]{mhchem}
\usepackage{physics}
\usepackage{enumitem}
\usepackage{mathrsfs}
\usepackage{mathtools}
\usepackage{subcaption}
\usetikzlibrary{arrows.meta,calc}

\usetikzlibrary{
  shapes.geometric,
  shadows,
  patterns,
  perspective,
  decorations,
  decorations.text
  }

\usetikzlibrary{decorations.markings,calc,decorations.pathreplacing}

\tikzset{
  baseline/.style = {line width=0.8pt, draw=black!70, line cap=round},
  tick/.style     = {line width=0.9pt, draw=black, line cap=round},
  site/.style     = {fill=black, draw=none},
  Jbond/.style    = {line width=1.4pt, draw=green!70!black, dotted, line cap=round},
  gbond/.style    = {line width=1.4pt, draw=cyan!80!black, line cap=round},
  Jlab/.style     = {font=\bfseries\footnotesize, text=green!60!black},
  glab/.style     = {font=\bfseries\footnotesize, text=cyan!80!black},
  upA/.style      = {-stealth, line width=0.5mm, draw=red!80!black},
  dnA/.style      = {-stealth, line width=0.5mm, draw=blue!80!black}
}

\newcommand{\nicechain}[4]{%
  \begin{scope}[shift={#1}]
    \def\dx{1.20}\def\Larr{1.00}\def\yoff{0.50}\def\rsite{0.12}\def\ticklen{0.17}
    \newcount\Np \Np=0
    \foreach \s [count=\j] in {#2} {\global\Np=\j}
    \newcount\N \N=\numexpr\Np-1\relax
    \ifnum\Np>1
      \draw[Jbond] ({1*\dx},0) -- ({2*\dx},0);
      \path ($({1*\dx},0)!0.5!({2*\dx},0)$) ++(0,0.20) node[Jlab]{J};
    \fi
    \ifnum\N>2
      \foreach \j in {2,...,\numexpr\N-1\relax}{
        \pgfmathtruncatemacro{\jp}{\j+1}
        \draw[gbond] ({\j*\dx},0) -- ({\jp*\dx},0);
        \path ($({\j*\dx},0)!0.5!({\jp*\dx},0)$) ++(0,0.20) node[glab]{g};
      }
    \fi
    \ifnum\Np>2
      \draw[Jbond] ({\N*\dx},0) -- ({\Np*\dx},0);
      \path ($({\N*\dx},0)!0.5!({\Np*\dx},0)$) ++(0,0.20) node[Jlab]{J};
    \fi
    \foreach \s [count=\j] in {#2}{
      \draw[tick] ({\j*\dx},-\ticklen) -- ({\j*\dx},\ticklen);
      \fill[site] ({\j*\dx},0) circle (\rsite);
      \ifnum\s=1
        \draw[upA] ({\j*\dx},-\yoff) -- ++(0,\Larr);
      \else
        \draw[dnA] ({\j*\dx},\yoff) -- ++(0,-\Larr);
      \fi
    }
  \end{scope}
}
\usepackage{xcolor}

\definecolor{linkcolor}{RGB}{0,0,255}      % blue
\definecolor{citecolor}{RGB}{0,128,0}     % green
\definecolor{urlcolor}{RGB}{255,0,0}      % red
\definecolor{lossred}{RGB}{211,47,47}
\definecolor{gainblue}{RGB}{25,118,210}
\definecolor{bulkdark}{RGB}{38,50,56}
\definecolor{latgray}{RGB}{150,158,165}
\definecolor{boxgray}{RGB}{90,98,105}
\definecolor{panelbg}{RGB}{248,249,251}

\usepackage{hyperref}
\hypersetup{
  colorlinks   = true,
  linkcolor    = linkcolor,
  citecolor    = citecolor,
  urlcolor     = urlcolor,
  linktoc      = all,
  pdfborder    = {0 0 0},
}

\definecolor{dy}{rgb}{0.9,0.9,0.4}
\definecolor{dr}{rgb}{0.95,0.65,0.55}
\definecolor{db}{rgb}{0.5,0.8,0.9}
\definecolor{dg}{rgb}{0.2,0.9,0.6}
\definecolor{BrickRed}{rgb}{0.8,0.3,0.3}
\definecolor{Navy}{rgb}{0.2,0.2,0.6}
\definecolor{DarkGreen}{rgb}{0.1,0.4,0.1}
\definecolor{phaseK}{HTML}{CC79A7}
\definecolor{phaseY}{HTML}{56B4E9}
\definecolor{phaseU}{HTML}{E69F00}
\definecolor{phaseF}{HTML}{23EB91}

\begin{document}

\title{ Monotonic and non-Nonmonotonic Impurity Entropy flows \\ in a $\mathscr{PT}$-Symmetric Kondo Model}
\author{Pradip Kattel}
\email{pradip.kattel@unige.ch}
\affiliation{Department of Quantum Matter Physics, University of Geneva, Quai Ernest-Ansermet 24, 1211 Geneva, Switzerland}
\author{Abay Zhakenov}
\author{Natan Andrei}
\affiliation{Department of Physics and Astronomy, Center for Materials Theory, Rutgers University, Piscataway, New Jersey 08854, USA
}

\begin{abstract}
Quantum impurity models provide a paradigmatic setting for studying Kondo screening, boundary criticality, and impurity entropy. While these phenomena are well understood in unitary systems, their fate in non-Hermitian many-body settings remains largely unexplored. We study a $\mathscr{PT}$-symmetric quantum impurity model consisting of a unitary SU(2)$_1$ Wess--Zumino--Witten bulk attached to two impurity spins through complex-conjugate Kondo couplings. Using an integrable lattice realization solved by the Bethe Ansatz and benchmarked against finite-temperature matrix-product-state calculations, we determine the impurity free energy, entropy, and the corresponding $g$-function. We identify three distinct impurity phases. In the Kondo-screened phase, the spectrum remains real and the impurity entropy decreases monotonically from $\ln 4$ in the ultraviolet to $0$. Remarkably, this monotonic flow persists despite the nonunitary boundary interaction, beyond the standard domain of the $g$-theorem. The Yu--Shiba--Rusinov phase instead exhibits spontaneous $\mathscr{PT}$-symmetry breaking and loss of boundary conformal invariance, while the local-moment phase displays cyclic renormalization-group flow, with identical impurity entropy and $g$-value in the ultraviolet and infrared limits.
\end{abstract}

\maketitle

A hallmark of quantum impurity physics is the screening of a localized degree of freedom by a many-body cloud, producing a nontrivial boundary renormalization-group flow between distinct boundary fixed points~\cite{wilson1975renormalization,andrei1983solution,tsvelick1983exact,affleck1995conformal}. The corresponding impurity free energy and the $g$-function characterizing this boundary RG flow are defined by
\begin{equation}
F_{\rm imp}(T)\equiv F_{\rm total}(T)-F_0(T)
=-T\ln g_{\rm imp}(T),
\label{defFimp}
\end{equation}
where $F_{\rm total}(T)$ and $F_0(T)$ are the free energies of the system with and without the impurities, respectively. The impurity entropy is then
\begin{equation}
S_{\rm imp}(T)
=-\frac{\partial F_{\rm imp}(T)}{\partial T}
=\left(1+T\frac{\partial}{\partial T}\right)\ln g_{\rm imp}(T).
\label{def-fn-lng}
\end{equation}
At conformal boundary fixed points, $g_{\rm imp}$ is scale independent and reduces to the universal Affleck--Ludwig boundary degeneracies $g_{\rm UV,IR}$. The $g$-theorem states that, for unitary boundary renormalization-group flows, $g(T)$, and hence the impurity entropy $S_{\rm imp}(T)$, decreases monotonically from the ultraviolet to the infrared~\cite{affleck1991universal,friedan2004boundary}. In the conventional spin-$\frac12$ Kondo problem, this corresponds to the screening of a free local moment, with the impurity entropy decreasing from $\ln2$ in the ultraviolet to $0$ at the strong-coupling screened fixed point.

Whether such ideas continue to hold in non-Hermitian quantum impurity systems remains largely unknown. Recent works have suggested that generalized notions of boundary and defect entropy can remain meaningful in non-unitary critical systems whenever the spectrum relevant to the partition function remains real~\cite{Liu2026ExtractingBC,sinha2026lattice,Maeda2026Holographic}. At the same time, non-Hermitian Kondo models have emerged as a particularly interesting setting in which to revisit these questions, motivated by their realization in dissipative quantum matter, open quantum systems, and ultracold atoms with inelastic scattering~\cite{nakagawa2018non,kattel2025dissipation,yi2026non,yi2026interplay}.
\begin{figure}[H]
    \centering
    \includegraphics[width=\linewidth]{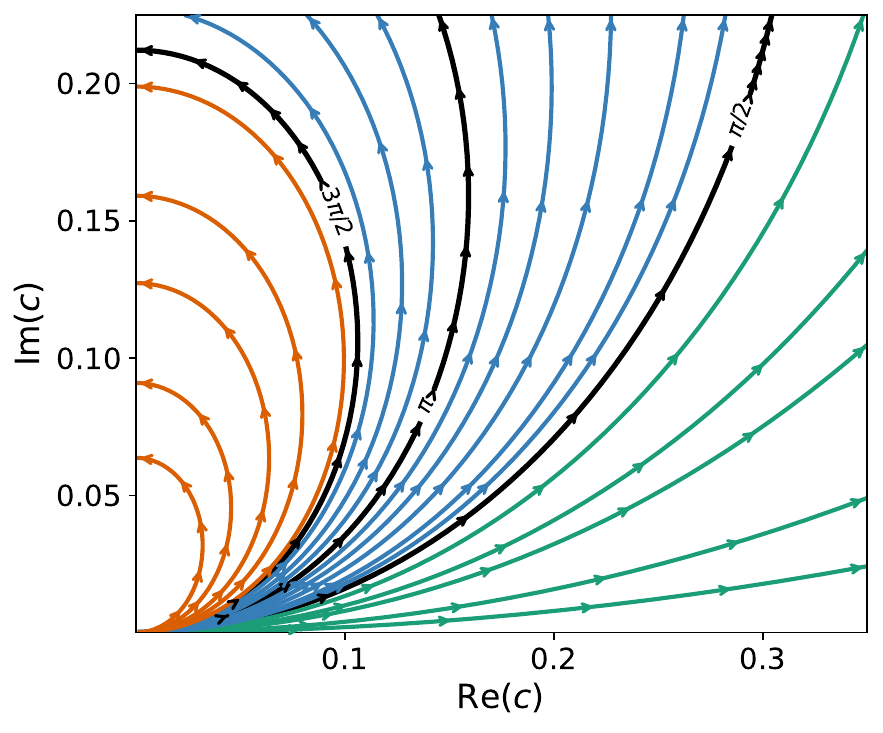}
    \caption{
Renormalization-group trajectories in the complex-coupling plane
$\tilde c = c_R + i c_I$. The solid black curves correspond to
$\alpha=\pi/2$, $\pi$, and $3\pi/2$, separating the Kondo regime
(green), two YSR subphases (blue), and the local-moment regime (orange).
The Kondo regime flows from weak to strong coupling, the YSR regime
is characterized by a bound state that generates a finite
energy scale, and the local-moment regime exhibits flow reversal
toward weak coupling. Arrows indicate the direction of the RG flow.
}
    \label{fig:rgflow}
\end{figure}

These studies have revealed that non-Hermitian Kondo models can exhibit renormalization-group behavior with no analog in conventional impurity physics. In particular, Ref.~\cite{nakagawa2018non} found perturbatively that non-Hermiticity can induce a reversion of RG flows and invalidate the conventional $g$-theorem. Subsequently, Ref.~\cite{kattel2025dissipation} confirmed this picture nonperturbatively via the Bethe Ansatz, obtaining the exact RG-invariant scale
$T_0=T_Ke^{i\alpha}$,
where $T_K=2D\exp[-\pi\cos\phi/c]$ and $\alpha=\pi\sin\phi/c$ for the complex Kondo coupling $\tilde c=ce^{i\phi}$. The corresponding RG trajectories are shown in Fig.~\ref{fig:rgflow}. Depending on the RG invariant $\alpha$, the flow exhibits three qualitatively distinct regimes: a Kondo phase flowing from weak to strong coupling, a Yu--Shiba--Rusinov (YSR) phase in which a boundary bound state appears  generating a finite energy scale and breaking scale invariance, and a local-moment phase in which the RG flow becomes cyclic. Such behavior has no counterpart in conventional unitary impurity systems and raises a fundamental question: how does the finite temperature impurity entropy behave across these distinct non-Hermitian RG regimes?

Among these three regimes, the Kondo and local-moment phases remain in the $\mathscr{PT}$-unbroken sector and possess entirely real spectra, while the intermediate YSR phase spontaneously breaks $\mathscr{PT}$ symmetry and develops complex-conjugate energy eigenvalues~\cite{kattel2025spin}. Consequently, equilibrium thermodynamics is well defined throughout the Kondo and local-moment phases but not in the YSR regime. The Kondo phase exhibits the conventional crossover from weak-coupling ultraviolet to strong-coupling infrared, terminating at a screened impurity fixed point despite the absence of Hermiticity~\cite{kattel2025dissipation,kattel2025spin,yi2026interplay}. By contrast, the local-moment phase is governed by cyclic renormalization-group flow, while the YSR phase is characterized by boundary bound states and the loss of boundary conformal invariance.

In this work, we investigate the impurity entropy throughout the entire $\mathscr{PT}$-unbroken phase diagram. To this end, we study an integrable $\mathscr{PT}$-symmetric extension of the non-Hermitian Kondo model in which two boundary impurities are coupled through complex-conjugate Kondo interactions. Upon non-Abelian bosonization~\cite{witten1984non}, the spin sector is described by the unitary $SU(2)_1$ Wess--Zumino--Witten theory, while the impurities appear as a pair of non-Hermitian defect lines related by $\mathscr{PT}$ symmetry. The resulting continuum Hamiltonian is $H=H_{\rm WZW}+H_{\rm imp}$ with
\begin{equation}
H_{\mathrm{imp}}
=
\lambda\left(\mathbf J_L+\mathbf J_R\right)_{x=0}\cdot\mathbf S_L
+
\lambda^*\left(\mathbf J_L+\mathbf J_R\right)_{x=L}\cdot\mathbf S_R.
\end{equation}
Thus, non-unitarity enters through the insertion of a pair of complex-valued defect lines at the boundaries. The central question is whether the associated impurity entropy remains a meaningful characterization of boundary renormalization-group flow despite the loss of unitarity, and how it reflects the distinct Kondo- and local-moment phases of the $\mathscr{PT}$-unbroken regime. Holographic realizations of related $\mathscr{PT}$-symmetric BCFTs with exactly marginal defect operators have recently been investigated in Ref.~\cite{Maeda2026Holographic}. In contrast, the boundary Kondo interaction considered here is marginally relevant and generates a genuine boundary RG flow.

We investigate a lattice regularization of the continuum problem, realized through an integrable $\mathscr{PT}$-symmetric spin chain. The microscopic Hamiltonian is
\begin{equation}
H = J\sum_{j=1}^{N-1} \boldsymbol{\sigma}_j\cdot\boldsymbol{\sigma}_{j+1} + J_{\mathrm{imp}}
\boldsymbol{\sigma}_1\cdot\boldsymbol{\sigma}_L
+J_{\mathrm{imp}}^* \boldsymbol{\sigma}_{N}\cdot\boldsymbol{\sigma}_R,
\end{equation}
where $\boldsymbol{\sigma}_L$ and $\boldsymbol{\sigma}_R$ are the two impurity spins at the left and right ends of the chain. We parametrize the complex boundary coupling as
$J_{\mathrm{imp}} = \frac{J}{1-\left(\beta+i\gamma\right)^2}.$
The complex-conjugate boundary couplings render the Hamiltonian $\mathscr{PT}$ symmetric, while $\gamma$ and $\beta$ control the degree of non-Hermiticity and the boundary phase structure, respectively.

\begin{figure}
\centering
\begin{tikzpicture}[
    >=Latex,
    spin/.style={circle,draw=bulkdark,fill=bulkdark,inner sep=0pt,minimum size=5.5pt},
    imp/.style={circle,draw=black!55,line width=0.5pt,inner sep=0pt,minimum size=11pt},
    upar/.style={-{Latex[length=3.6pt,width=3.6pt]},line width=0.9pt,bulkdark},
    cuptag/.style={fill=white,inner sep=1.2pt,rounded corners=1pt,font=\footnotesize},
    font=\small
]
\def\dx{1.05}\def\y{0}\def\R{5.25}\def\amp{0.20}
 
% optical-lattice potential
\draw[latgray,line width=0.9pt,samples=141,domain=-0.45:5.7]
      plot (\x,{-\amp*cos((\x)/1.05*360)});
 
% bulk exchange bonds (white halo cuts lattice)
\foreach \i in {1,...,3}{
  \pgfmathsetmacro\xa{\i*\dx}\pgfmathsetmacro\xb{(\i+1)*\dx}
  \draw[white,line width=2.8pt] (\xa,\y)--(\xb,\y);
  \draw[bulkdark,line width=1.2pt] (\xa,\y)--(\xb,\y);}
 
% boundary (impurity) bonds: PT-conjugate pair
\draw[white,line width=3.0pt] (0,\y)--(\dx,\y);
\draw[white,line width=3.0pt] (4*\dx,\y)--(\R,\y);
\draw[lossred ,line width=1.4pt,dash pattern=on 3.5pt off 2pt] (0,\y)--(\dx,\y);
\draw[gainblue,line width=1.4pt,dash pattern=on 3.5pt off 2pt] (4*\dx,\y)--(\R,\y);
 
% optical tweezers
\foreach \cx/\cc in {0/lossred, \R/gainblue}{
  \fill[\cc,opacity=0.13] (\cx,\y+0.85)--(\cx-0.40,\y+1.95)--(\cx+0.40,\y+1.95)--cycle;
  \draw[\cc,opacity=0.55,line width=0.55pt] (\cx-0.40,\y+1.95)--(\cx,\y+0.85)--(\cx+0.40,\y+1.95);}
 
% bulk spins (alternating up/down)
\foreach \i in {1,...,4}{
  \pgfmathsetmacro\xx{\i*\dx}\node[spin] (s\i) at (\xx,\y) {};
  \pgfmathparse{int(mod(\i,2))}
  \ifnum\pgfmathresult=1 \draw[upar] (\xx,\y-0.26)--(\xx,\y+0.32);
  \else \draw[upar] (\xx,\y+0.26)--(\xx,\y-0.32);\fi}
 
% impurity spins
\node[imp,fill=lossred]  (L)  at (0,\y) {};
\draw[->,white,line width=0.9pt] (0,\y+0.28)--(0,\y-0.28);
\node[imp,fill=gainblue] (Rr) at (\R,\y) {};
\draw[->,white,line width=0.9pt] (\R,\y-0.28)--(\R,\y+0.28);
 
% reservoirs (terse labels)
\draw[lossred ,line width=1.4pt,->] (0,\y-0.42)--(0,\y-1.05)
      node[below,lossred,align=center,font=\scriptsize\bfseries]
      {loss bath\\[-1pt]\mdseries$-i\gamma$};
\draw[gainblue,line width=1.4pt,<-] (\R,\y-1.05)--(\R,\y-0.42);
\node[below,gainblue,align=center,font=\scriptsize\bfseries] at (\R,\y-1.05)
      {gain bath\\[-1pt]\mdseries$+i\gamma$};
 
% spin labels
\node[lossred ,font=\bfseries\footnotesize] at (0,\y+0.58)  {$\boldsymbol{\sigma}_L$};
\node[gainblue,font=\bfseries\footnotesize] at (\R,\y+0.58) {$\boldsymbol{\sigma}_R$};
\node[bulkdark,font=\footnotesize] at (1*\dx,\y+0.60) {$\boldsymbol{\sigma}_1$};
\node[bulkdark,font=\footnotesize] at (4*\dx,\y+0.60) {$\boldsymbol{\sigma}_N$};
 
% tweezer titles (terse)
\node[lossred ,font=\scriptsize\bfseries] at (0,\y+2.18)  {tweezer $L$};
\node[gainblue,font=\scriptsize\bfseries] at (\R,\y+2.18) {tweezer $R$};
 
% coupling tags
\node[cuptag,text=lossred ,font=\scriptsize] at (0.5*\dx,\y-0.48) {$J_{\mathrm{imp}}$};
\node[cuptag,text=gainblue,font=\scriptsize] at (4.5*\dx,\y-0.48) {$J_{\mathrm{imp}}^{*}$};
\node[cuptag,text=bulkdark,font=\scriptsize] at (2.5*\dx,\y-0.48) {$J$};
 
% Hamiltonian panel (two lines)
\node[draw=boxgray,line width=0.6pt,rounded corners=3pt,fill=panelbg,
      inner xsep=8pt,inner ysep=5pt,align=center] (H) at (\R/2,\y+3.4)
   {$\begin{aligned}
       H=\;&J\sum_{j=1}^{N-1}\boldsymbol{\sigma}_j\!\cdot\!\boldsymbol{\sigma}_{j+1}+\color{lossred} J_{\mathrm{imp}}\,\boldsymbol{\sigma}_1\!\cdot\!\boldsymbol{\sigma}_L
            \color{black} + \color{gainblue} J_{\mathrm{imp}}^{*}\,\boldsymbol{\sigma}_N\!\cdot\!\boldsymbol{\sigma}_R
     \end{aligned}$};
 
\end{tikzpicture}
    \caption{Schematic realization of the $\mathscr{PT}$-symmetric spin chain. A one-dimensional optical lattice realizes the bulk Heisenberg chain, while site-selective optical tweezers locally control the boundary impurity spins. Coupling the impurities to engineered reservoirs generates complex-conjugate boundary exchange couplings $J_{\mathrm{imp}}$ and $J_{\mathrm{imp}}^{*}$, realizing a $\mathscr{PT}$-symmetric boundary deformation of an otherwise unitary spin chain.}
    \label{fig:experiment}
\end{figure}
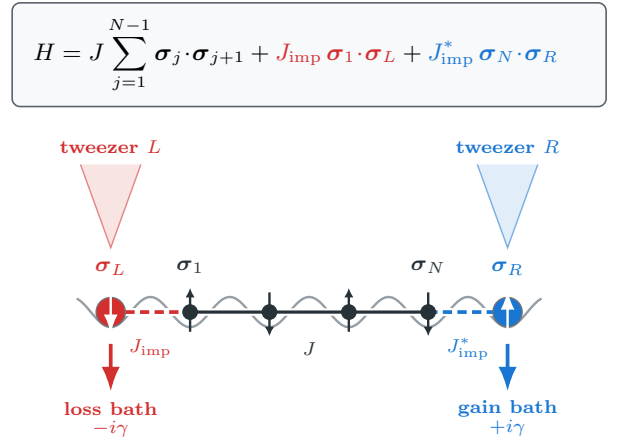

The lattice formulation serves several purposes. Besides providing an integrable regularization of the continuum impurity theory and enabling finite-temperature tensor-network calculations, it also suggests a possible route toward experimental realization, illustrated schematically in Fig.~\ref{fig:experiment}. In an ultracold-atom implementation, a one-dimensional optical lattice realizes the bulk spin chain, while site-selective optical tweezers locally control the boundary impurity spins. Coupling the impurities to engineered reservoirs can generate complex boundary exchange interactions, extending proposals for non-Hermitian spin dynamics in alkaline-earth systems~\cite{nakagawa2018non}. More broadly, recent advances in the control of dissipation, non-Hermitian dynamics, and $\mathscr{PT}$ symmetry in atomic, photonic, and Rydberg platforms~\cite{li2019observation,peng2014parity,feng2014single,hodaei2014parity,yamamoto2019theory,miri2019exceptional} suggest that the ingredients required to engineer complex-conjugate boundary couplings are increasingly accessible. We therefore complement the exact Bethe Ansatz solution with independent tensor-network simulations, providing a nonperturbative benchmark of the impurity thermodynamics.

The spectrum and phase diagram of this model were analyzed in Ref.~\cite{kattel2025spin}. As a function of $\beta$, the model exhibits four distinct boundary regimes: a Kondo-screened phase for $0<\beta<\frac12$, a bound-mode phase I for $\frac12<\beta<1$, a bound-mode phase II for $1<\beta<\frac32$, and an effectively unscreened local-moment phase for $\beta>\frac32$. The Kondo and local-moment phases remain in the $\mathscr{PT}$-unbroken regime and possess entirely real spectra, whereas the two bound-mode phases exhibit spontaneous $\mathscr{PT}$-symmetry breaking and complex-conjugate energy eigenvalues.

Here, we investigate the impurity entropy throughout the entire $\mathscr{PT}$-unbroken phase diagram, comprising the Kondo phase ($0<\beta<\frac12$) and the local-moment phase ($\beta>\frac32$). In both regimes, the spectrum remains entirely real and the model admits a well-defined equilibrium thermodynamics. By contrast, the intermediate bound-mode phases spontaneously break $\mathscr{PT}$ symmetry and develop complex eigenvalues, placing them outside the thermodynamic framework considered here.

\begin{figure}
    \centering    \includegraphics[width=\linewidth]{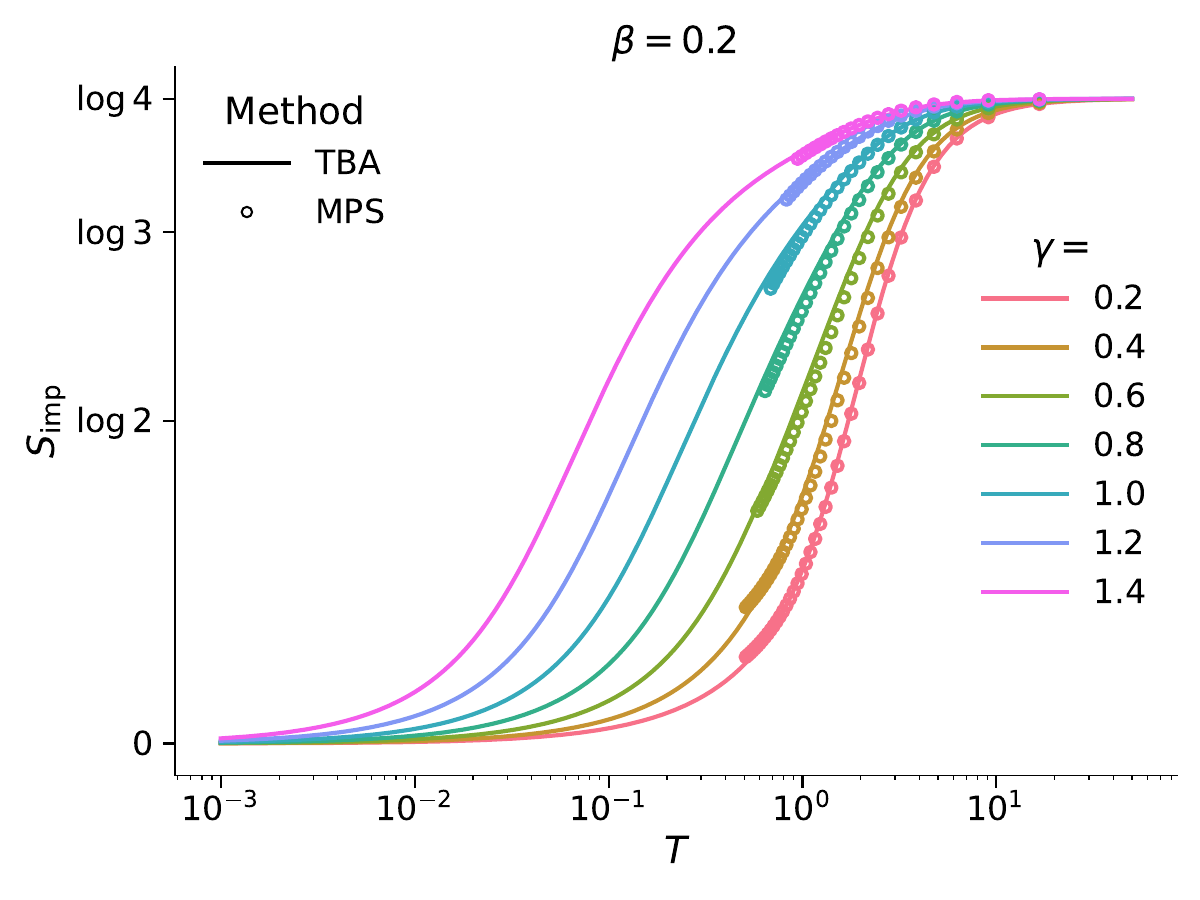}
    \caption{Impurity entropy $S_{\rm imp}$ as a function of temperature for $\beta=0.2$ and several values of the non-Hermiticity parameter
$\gamma$. Solid lines show the exact TBA solution, while the hollow circular markers denote finite-temperature matrix-product-state
(MPS) calculations. The entropy decreases monotonically from the ultraviolet value $\ln 4$ to the infrared value $0$. The excellent agreement between TBA and MPS provides an independent nonperturbative verification of the exact
thermodynamic solution.}
    \label{fig:entropy}
\end{figure}

Using the exact Bethe Ansatz solution, we compute the impurity contribution to the free energy and the corresponding impurity entropy $S_{\rm imp}$ associated with the pair of $\mathscr{PT}$-conjugate impurities throughout the $\mathscr{PT}$-unbroken regime. We find qualitatively distinct behaviors in the two phases. In the Kondo regime, the impurity entropy decreases monotonically from $\ln 4$ in the ultraviolet, where the two spin-$\frac{1}{2}$ impurities are effectively free, to $0$ in the infrared, where both impurities are screened (see Fig.~\ref{fig:entropy}). In the local-moment regime, by contrast, the cyclic renormalization-group flow is accompanied by a nonmonotonic impurity entropy, whose ultraviolet and infrared limits coincide. The impurity entropy therefore directly reflects the qualitatively different RG structures of the two regimes, including the unconventional flow reversal and cyclic behavior known to arise in non-Hermitian Kondo models~\cite{nakagawa2018non,kattel2025dissipation}.

The exact solution is obtained from the Bethe Ansatz equations
\begin{align}
&\left(\frac{\mu_j-\frac{i}{2}}{\mu_j+\frac{i}{2}}\right)^{2N}\prod_{\upsilon=\pm}
\frac{\mu_j+\upsilon (\gamma+i\beta)-\frac{i}{2}}{\mu_j+\upsilon (\gamma+i\beta)+\frac{i}{2}}
\frac{\mu_j+\upsilon (\gamma-i\beta)-\frac{i}{2}}{\mu_j+\upsilon (\gamma-i\beta)+\frac{i}{2}}\nonumber\\
&
\qquad\qquad=\prod_{l\neq j}^{M}\prod_{\upsilon=\pm}
\frac{(\mu_j+\upsilon\mu_l-i)}{(\mu_j+\upsilon\mu_l+i)}.
\end{align}
The energy and momentum of an eigenstate are determined by the corresponding Bethe root configuration $\{\mu_j,\, j=1,\ldots,M\}$. In particular, the energy takes the form
\begin{equation}
E=-\sum_{j=1}^{M}\frac{2J}{\mu_j^2+\frac14} +J(L-1)+\frac{2J(\gamma^2-\beta^2+1)} {\beta^4+2\beta^2(\gamma^2-1)+(\gamma^2+1)^2}.
\end{equation}
In the thermodynamic limit $L\to\infty$, the Bethe roots organize into continuous distributions of strings and holes. In the Kondo phase ($0<\beta<\frac12$), all eigenvalues remain real~\cite{kattel2025spin}. The corresponding finite-size Hamiltonian therefore admits a complete biorthogonal eigenbasis and a quasi-Hermitian description~\cite{mostafazadeh2001pseudo,mostafazadeh2002pseudo}, providing a natural framework for thermal equilibrium in the non-Hermitian setting~\cite{Lan2026KuboMartinSchwingerCF}. We may therefore introduce biorthogonal left and right eigenstates,
\begin{equation}
H|R_n\rangle = E_n|R_n\rangle, \qquad \langle L_n|H = E_n\langle L_n|,
\end{equation}
with $\langle L_n|R_m\rangle=\delta_{nm}$ and $\sum_n |R_n\rangle\langle L_n|=\mathbf 1$. Using the completeness relation above, the basis-independent trace may be evaluated in the biorthogonal eigenbasis. Since all eigenvalues remain real in the $\mathscr{PT}$-unbroken phase, the thermal partition function is therefore well defined and can be written as
\begin{equation}
Z = \mathrm{Tr}\,e^{-H/T} =\sum_n e^{-E_n/T}.
\end{equation}

Introducing the densities $\sigma_n(\lambda)$ and hole densities $\sigma_n^h(\lambda)$ associated with $n$-strings, the thermodynamic Bethe equations take the form
\begin{equation}
\sigma_n^h(\lambda) = f_n^{\mathrm{str}}(\lambda) - \sum_{m=1}^{\infty}
A_{nm}\sigma_m(\lambda).
\end{equation}
The free energy is obtained from the Yang--Yang functional~\cite{yang1969thermodynamics} $F=E-hS^z-T\mathcal S$,
where $\mathcal S$ is the Yang--Yang entropy. Minimizing the free energy with respect to the densities and introducing $\eta_n(\lambda)=\frac{\sigma_n^h(\lambda)}{\sigma_n(\lambda)}$, yields the thermodynamic Bethe Ansatz equations
\begin{equation}
\ln \eta_n(\lambda)= -\frac{2\pi}{T}\frac{\delta_{n,1}} {\cosh(\pi\lambda)}+G\ln\!\left[1+\eta_{n+1}\right]+ G\ln\!\left[1+\eta_{n-1}\right],
\end{equation}
where $Gf(\lambda)=\int d\mu\,\frac{f(\mu)} {2\cosh\!\left[\pi(\lambda-\mu)\right]}$.
The hierarchy is supplemented by the boundary conditions $\eta_0(\lambda)=0$, and
\begin{equation}
\lim_{n\to\infty}\left\{[n+1]\ln\left(1+\eta_n\right)-[n]\ln\left(1+ \eta_{n+1}\right)\right\}=-\frac{h}{T},
\end{equation}
where $h$ denotes an external magnetic field.

For an open chain, the free energy contains an extensive bulk contribution
as well as $O(1)$ boundary terms. In the definition of $F_{\rm imp}$ in
Eq.~\eqref{defFimp}, subtraction of the impurity-free free energy $F_0$ removes both
the extensive bulk contribution and the $O(1)$ contribution associated
with the universal free-boundary $g$-function of the open chain
~\cite{dorey2004integrable,dorey2006reflection,pozsgay2010mathcal,he2024exact},
thereby isolating the impurity contribution. Notice that, as in the conventional Kondo problem, the TBA equations themselves are independent of the impurity couplings. The impurity parameters enter only through the $O(1)$ boundary contribution to the free energy. Since the two boundary couplings are related by $\mathscr{PT}$ conjugation, the physical impurity free energy is obtained from the combined contribution of the two boundaries $F^{\mathrm{imp}}=F^{\mathrm{imp}}_L+F^{\mathrm{imp}}_R$. The evaluation of the impurity free energy is identical to that of the Hermitian boundary Kondo problem; we therefore omit the intermediate steps and refer the reader to the Supplementary Materials of Ref.~\cite{zhakenov2025thermodynamics}. In the Kondo-screened phase, where the impurity spectrum is organized into a
single thermodynamic tower, the resulting impurity free energy is
\begin{equation}
F_{\rm imp}=\frac{-T}{2}\sum_{\upsilon=\pm}\int d\lambda\,
\frac{\cosh\left[\pi\left(\lambda+\upsilon\gamma\right)\right]
\cos(\pi\beta)\ln\left(1+\eta_1(\lambda)\right)}{\cosh^2\left[\pi\left(\lambda+\upsilon\gamma\right)\right]-
\sin^2(\pi\beta)}.
\end{equation}

The ultraviolet and infrared limits of the impurity entropy can be obtained analytically. In the high-temperature limit, the driving term in the TBA equations vanishes, and the hierarchy admits the constant solution $\eta_n=n(n+2)$. Substituting this solution into the impurity free energy and using Eq.\eqref{def-fn-lng} yields
\begin{equation}
S_{\rm imp}(T\to\infty)=\ln 4,
\end{equation}
corresponding to two free spin-$\frac12$ impurities.

In the opposite limit $T\to0$, the asymptotic solution of the TBA equations gives
\begin{equation}
F_{\rm imp} = -\frac{T^2}{12} \cos(\pi\beta)\cosh(\pi\gamma) + O(T^3),
\end{equation}
from which it follows that $S_{\rm imp}(T\to0)=0$. Thus, despite the non-Hermitian boundary coupling, both impurities are completely screened in the infrared. Matching the low-temperature free energy to the universal local Fermi-liquid form $F_{\rm imp} = -\frac{\pi^2T^2}{6T_K}$   \cite{andrei1981scales, hewson1997kondo,andrei1983solution},
we obtain the Kondo scale
$T_K = \frac{2\pi^2J}{\cos(\pi\beta)\cosh(\pi\gamma)}$. Note that increasing $\gamma$ suppresses the Kondo scale exponentially,
$T_K\sim e^{-\pi\gamma}$ for $\gamma\gg1$. The impurity entropy $S_{\rm imp}$ therefore interpolates between the ultraviolet value $\ln4$ and the infrared value $0$, with the crossover governed by $T_K$. The crucial remaining question is whether the impurity entropy remains monotonic throughout the crossover, despite the absence of the assumptions underlying the Affleck--Ludwig $g$-theorem. In fact, the monotonicity can be established analytically by arguments given in \cite{andrei1983solution}, but to explicitly determine the full crossover between the ultraviolet and infrared fixed points, it is necessary to solve the TBA equations numerically. 
To benchmark the exact thermodynamic solution, we performed independent finite-temperature matrix-product-state calculations using the ITensor library~\cite{fishman2022itensor,fishman2022codebase}. Starting from the infinite-temperature density matrix $\rho \propto \mathbf{1}$, thermal states were generated by direct imaginary-time evolution of $\rho$ as a matrix-product operator. The impurity entropy was then obtained by subtracting the entropy of the corresponding impurity-free chain ($N=100$) from that of the full system ($N=102$).
Fig~\ref{fig:entropy} compares the exact TBA result with MPS calculations for a representative value of $\beta=0.2$ and several values of the non-Hermiticity parameter $\gamma$. For all parameters shown, the impurity entropy $S_{\rm imp}$ decreases monotonically from the ultraviolet value $\ln 4$ to the infrared value $0$, with no indication of non-monotonic behavior throughout the crossover. Moreover, numerical solutions of the TBA equations for other values of $\beta$ within the Kondo phase exhibit the same monotonic flow. Thus, throughout the $\mathscr{PT}$-symmetric Kondo regime, the impurity entropy remains monotonic despite the absence of boundary unitarity and the unconventional RG behavior characteristic of non-Hermitian Kondo systems.

\begin{figure}[H]
    \centering
    \includegraphics[width=\linewidth]{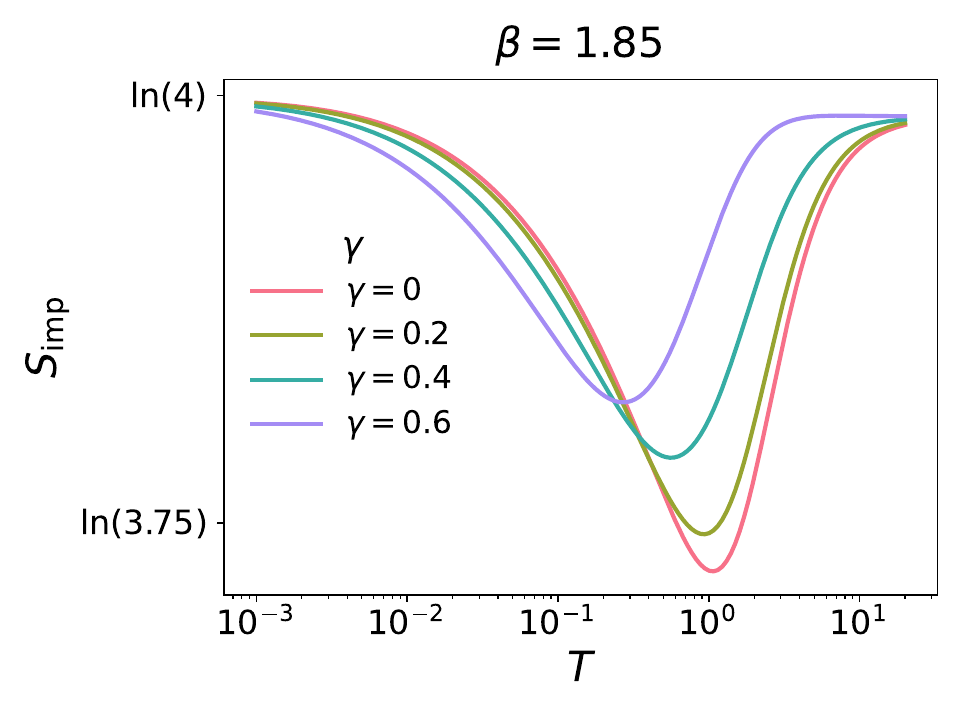}
    \caption{In the local-moment phase, the cyclic boundary RG flow yields the
same impurity entropy in the ultraviolet and infrared,
$S_{\rm imp}^{\rm UV}=S_{\rm imp}^{\rm IR}=\ln 4$, while $S_{\rm imp}$
 dips below the UV value at intermediate temperatures.}
    \label{fig:LM}
\end{figure}

In the local-moment phase, the impurity contribution is obtained from the
three-tower construction~\cite{zhakenov2025thermodynamics}. Writing
\begin{equation}
n=\lfloor2\beta\rfloor,\qquad
\epsilon=2\beta-n,\qquad
\bar\epsilon=1-\epsilon,
\end{equation}
$L_m(\lambda)=\ln[1+\eta_m(\lambda)]$ and
\begin{equation}
K_c^{(\gamma)}(\lambda)
=
\sum_{\nu=\pm}
\frac{1}{
\cosh\!\left[
\pi\left(\lambda+\frac{i\nu}{2}(c+i\gamma)\right)
\right]},
\end{equation}
the three tower free energies of the left impurity are compactly
\begin{equation}
\begin{pmatrix}
F_{L}^{\mathcal T_1}\\
F_{L}^{\mathcal T_2}\\
F_{L}^{\mathcal T_3}
\end{pmatrix}
=
\frac{T}{4}\int d\lambda
\begin{pmatrix}
-K_\epsilon^{(\gamma)} & K_{\bar\epsilon}^{(\gamma)} & 0 & 0\\
0&0&K_\epsilon^{(\gamma)}&-K_{\bar\epsilon}^{(\gamma)}\\
0&K_{\bar\epsilon}^{(\gamma)}&K_\epsilon^{(\gamma)}&0
\end{pmatrix}
\begin{pmatrix}
L_{n+1}\\L_n\\L_{n-1}\\L_{n-2}
\end{pmatrix}.
\label{eq:LM-towers}
\end{equation}
The right-impurity contributions follow by $\mathscr{PT}$ conjugation,
$F_{R}^{\mathcal T_a}
=(F_{L}^{\mathcal T_a})^*$.
Hence, the total free energy of the two impurities is
\begin{equation}
F_{\rm imp}(T)
=
-2T\,\Re\ln\!\left[
\sum_{a=1}^{3}
e^{-F_{L}^{\mathcal T_a}(T)/T}
\right].
\label{eq:LM-Fimp}
\end{equation}
As shown in Fig.~\ref{fig:LM}, the impurity entropy approaches $\ln 4$
in both the UV and IR limits, consistent with the cyclic boundary RG flow
of the local-moment phase illustrated in Fig.~\ref{fig:rgflow}.

Our results demonstrate that boundary unitarity is not a prerequisite for monotonic impurity entropy in the Kondo-screened phase. For the representative case shown in Fig.~\ref{fig:entropy}, the agreement between the exact TBA solution and independent finite-temperature MPS calculations is excellent across the entire crossover regime, providing a nonperturbative verification of the thermodynamics. In contrast, in the local-moment phase the impurity entropy exhibits a nonmonotonic temperature dependence and approaches $\ln 4$ in both the UV and IR limits, consistent with the cyclic nature of the boundary RG flow.

In this work, we investigated impurity entropy in an integrable $\mathscr{PT}$-symmetric non-Hermitian Kondo model. Using the exact Bethe Ansatz solution, supplemented by finite-temperature matrix-product-state calculations, we determined the impurity free energy and entropy throughout the $\mathscr{PT}$-unbroken Kondo-screened and local-moment phases. In the Kondo-screened phase, the impurity entropy $S_{\rm imp}$ decreases monotonically from the ultraviolet value $\ln 4$ to the infrared value $0$, corresponding to the complete screening of the two impurity spins. In the local-moment phase, by contrast, $S_{\rm imp}$ is nonmonotonic at intermediate temperatures but approaches $\ln 4$ in both the ultraviolet and infrared limits, reflecting the cyclic boundary RG flow.

The persistence of monotonicity in the Kondo phase is particularly striking because the boundary perturbation is non-Hermitian and lies outside the class of boundary RG flows for which the Affleck--Ludwig $g$-theorem has been established. Nevertheless, throughout the $\mathscr{PT}$-symmetric Kondo phase, the impurity entropy remains a monotonic measure of the impurity degrees of freedom, with the low-temperature thermodynamics governed by a single Kondo scale $T_K$. This behavior contrasts sharply with the local-moment phase, where the cyclic boundary RG flow leads to a nonmonotonic impurity entropy that approaches the same value, $\ln 4$, in both the UV and IR limits.

Our results demonstrate that key signatures of Kondo screening and boundary criticality persist well beyond the context of unitary quantum impurity systems. In particular, boundary unitarity is not essential for monotonic impurity entropy in the Kondo-screened phase, where the boundary RG flow remains irreversible despite the non-Hermitian interaction. By contrast, the local-moment phase exhibits a cyclic boundary RG flow accompanied by a nonmonotonic impurity entropy. Since renormalization-group irreversibility has been rigorously established for broad classes of non-unitary but $\mathscr{PT}$-symmetric quantum field theories through a generalization~\cite{castro2017irreversibility} of Zamolodchikov's $c$-theorem~\cite{zamolodchikov1986irreversibility}, our exact results suggest that an analogous extension of the Affleck--Ludwig $g$-theorem~\cite{affleck1991universal,friedan2004boundary,casini2016g} may hold within the $\mathscr{PT}$-symmetric Kondo phase. Establishing the precise conditions for such a theorem would place the monotonic impurity entropy and irreversible boundary RG flow observed in the Kondo phase on a rigorous 
and at the same time clarify its failure in the local moment phase.

\textit{Acknowledgments:} We thank Parameshwar R. Pasnoori, Patrick Azaria, and J. H. Pixley for prior collaborations on related topics, and Yicheng Tang for useful discussions. This work was supported by the Swiss National Science Foundation under Division II (Grant No.~200020-219400).
\bibliography{ref}
\end{document}